\documentclass{ar-1col-S2O}
\usepackage[numbers]{natbib}
\usepackage{url}
\usepackage{rotating} 
\usepackage{amsmath}  
\usepackage{lineno}   
\graphicspath{{./figures/}}

\jname{Xxxx. Xxx. Xxx. Xxx.}
\jvol{AA}
\jyear{YYYY}
\doi{10.1146/((please add article doi))}

\begin{document}


\markboth{J. Heise}{Construction and Science of SURF}

\title{Construction and Science of SURF}

\author{Jaret Heise
\affil{South Dakota Science and Technology Authority, Sanford Underground Research Facility, Lead, USA, 57754-1700; email: jheise@sanfordlab.org}
}

\begin{abstract}
The Sanford Underground Research Facility (SURF) began operation in 2007 as a facility dedicated to advancing compelling multidisciplinary scientific research. SURF is one of the deepest laboratory sites and offers the largest footprint in the world for scientific pursuits, including physics campuses on the 4850-foot level where the LUX-ZEPLIN, {\sc Majorana Demonstrator}, and CASPAR experiments are located. SURF is also home to the Long-Baseline Neutrino Facility (LBNF) that will host the international Deep Underground Neutrino Experiment (DUNE). SURF provides ultra-low background environments, low-background assay capabilities, and electroformed copper is produced at the facility. In this review, we discuss the history, features and status of the facility, as well as the current scientific program and future evolution and plans.
\end{abstract}

\begin{keywords}
underground laboratory, dark matter, neutrinos, double-beta decay, nuclear astrophysics, quantum computing and sensors
\end{keywords}
\maketitle

\tableofcontents


\section{INTRODUCTION}

\begin{figure}[!htbp]
    \centering
    \includegraphics[width=\textwidth]{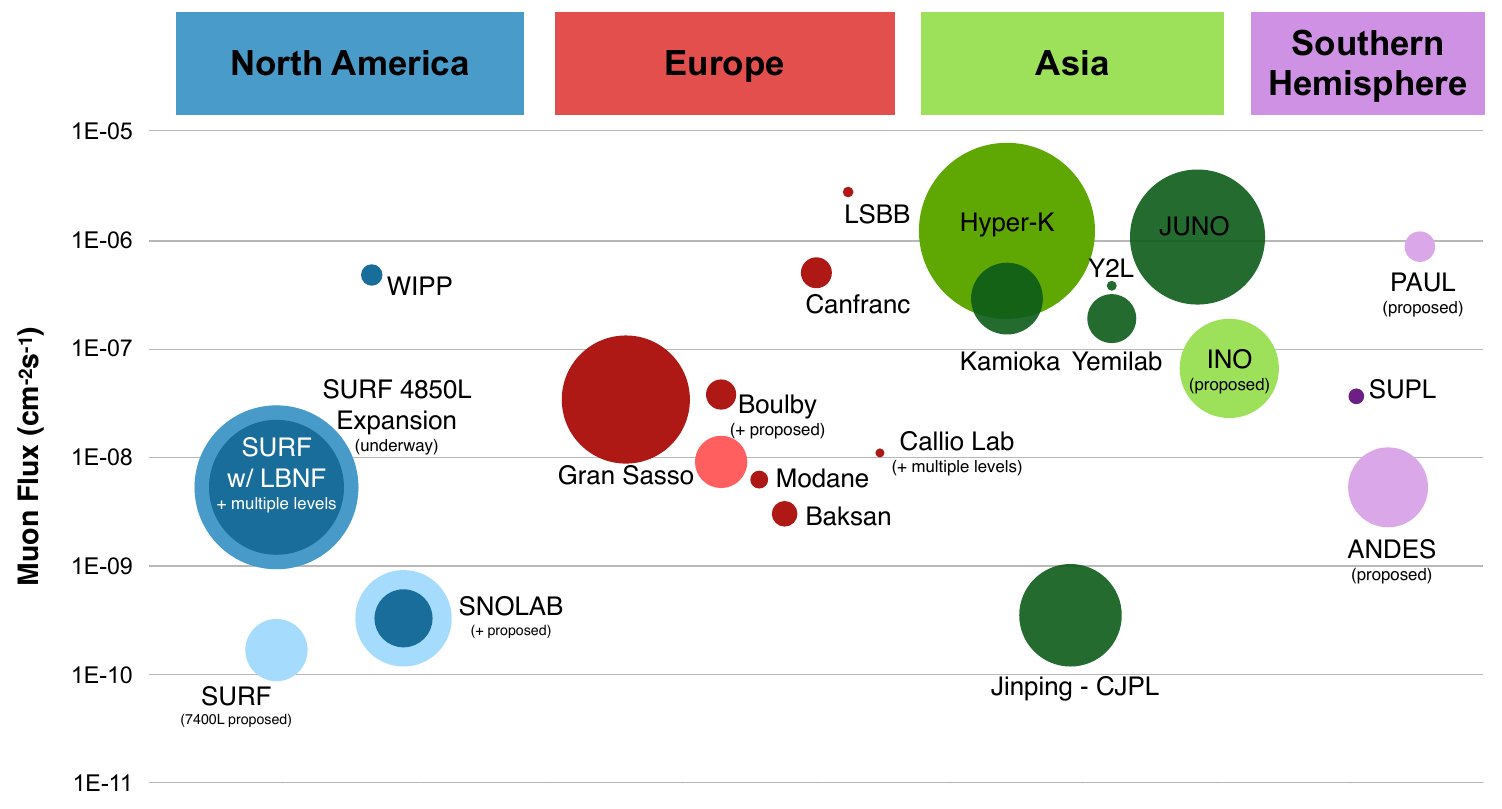}
    \caption{The size (volume of science space) and effective shielding depth (total muon flux) for the main global deep underground facilities are represented according to geographic location. The far left-hand side shows the current state and future of SURF, where the dark-blue circle represents the current combined size of 4850L laboratories. The SURF strategic plan aims to provide additional lab space on the 4850L as well as the possibility of a deeper site on the 7400L as indicated by lighter shaded blue circles. Some muon-flux values are estimated using a recent parameterization~\cite{JNE:2020bwn}.}
    \label{fig:LabVolumes}
\end{figure}

The Sanford Underground Research Facility (SURF) is an international facility dedicated to advancing compelling multidisciplinary underground scientific research, including physics, biology, geology and engineering~\cite{Heise:2023tln, Heise:2022iaf, Heise:2014gta, Lesko:2012fp, Heise:2010juv, Alonso:2009zz}. SURF's vast underground environment (see Figure~\ref{fig:LabVolumes}), allows researchers to explore an array of important questions regarding the origin of life and its diversity, mechanisms associated with geologic processes as well as a number of engineering topics such as mining innovations and technology developments. A deep underground laboratory is also where some of the most fundamental topics in physics can be investigated, including the nature of dark matter, the properties of neutrinos and topics related to nuclear astrophysics such as the synthesis of atomic elements within stars. In addition to advancing world-class science, SURF's mission also includes inspiring learning across generations~\cite{HornWoodward:2023}.
\begin{marginnote}[]
\entry{SURF}{Sanford Underground Research Facility}
\end{marginnote}

With strong support from the scientific community as well as federal, state and private (T.\ Denny Sanford) funding, SURF has been operating as a dedicated research facility for over 18 years. Initial concepts for SURF were developed with the support of the U.S.\ National Science Foundation (NSF) as the primary site for the Deep Underground Science and Engineering Laboratory (DUSEL)~\cite{Lesko:2011qk}. Since Fall 2011, SURF operation has been funded by the U.S.\ Department of Energy's (DOE) Office of Science, initially via sub-contracts with various national laboratories and through a Cooperative Agreement since Fall 2019. Cooperative Agreement funding is approved in increments, and the second five-year term started in Fall 2024.
\begin{marginnote}[]
\entry{NSF}{National Science Foundation}
\entry{DUSEL}{Deep Underground Science and Engineering Laboratory}
\entry{DOE}{Department of Energy}
\end{marginnote}

SURF is located in Lead, SD at the site of the former Homestake Gold Mine, which was the largest and deepest gold mine in the western hemisphere. The SURF site is renowned for its unique role in the pioneering collaboration between John Bahcall and Ray Davis and the chlorine-based solar neutrino experiment constructed on the 4850-foot level (4850L) of the mine starting in the mid-1960s~\cite{Bahcall:1964gx, Davis:1964hf}~\footnote{Based on the importance of this historical connection, SURF was designated an American Physical Society Historical Site on September 2020, with a dedication ceremony, including members of the Davis family, held in May 2022~\cite{APS:2022historical}.}. Early underground dark matter measurements were also made at the Homestake site~\cite{Ahlen:1987mn}. The extensive infrastructure of the former mine has since been prized for continued scientific use. When Barrick Gold Corporation ceased operation of the Homestake Mine in 2003, the scientific community seized the opportunity to convert the mine into a dedicated underground research facility, and Homestake was selected by the NSF as the DUSEL site in July 2007.

Since 2004, the South Dakota Science and Technology Authority (SDSTA) has provided management and administrative support necessary to host world-leading science at SURF, including a process for evaluating research proposals and implementing experiments~\cite{SURF_science}. The SDSTA organization currently comprises 220 full/part-time staff in 9 departments, 5 offices, and including the Institute for Underground Science at SURF, the Sanford Lab Homestake Visitor Center and the SURF Foundation.
\begin{marginnote}[]
\entry{SDSTA}{South Dakota Science and Technology Authority}
\end{marginnote}



\section{FACILITY}

\begin{figure}[!htbp]
    \centering
    \includegraphics[width=\textwidth]{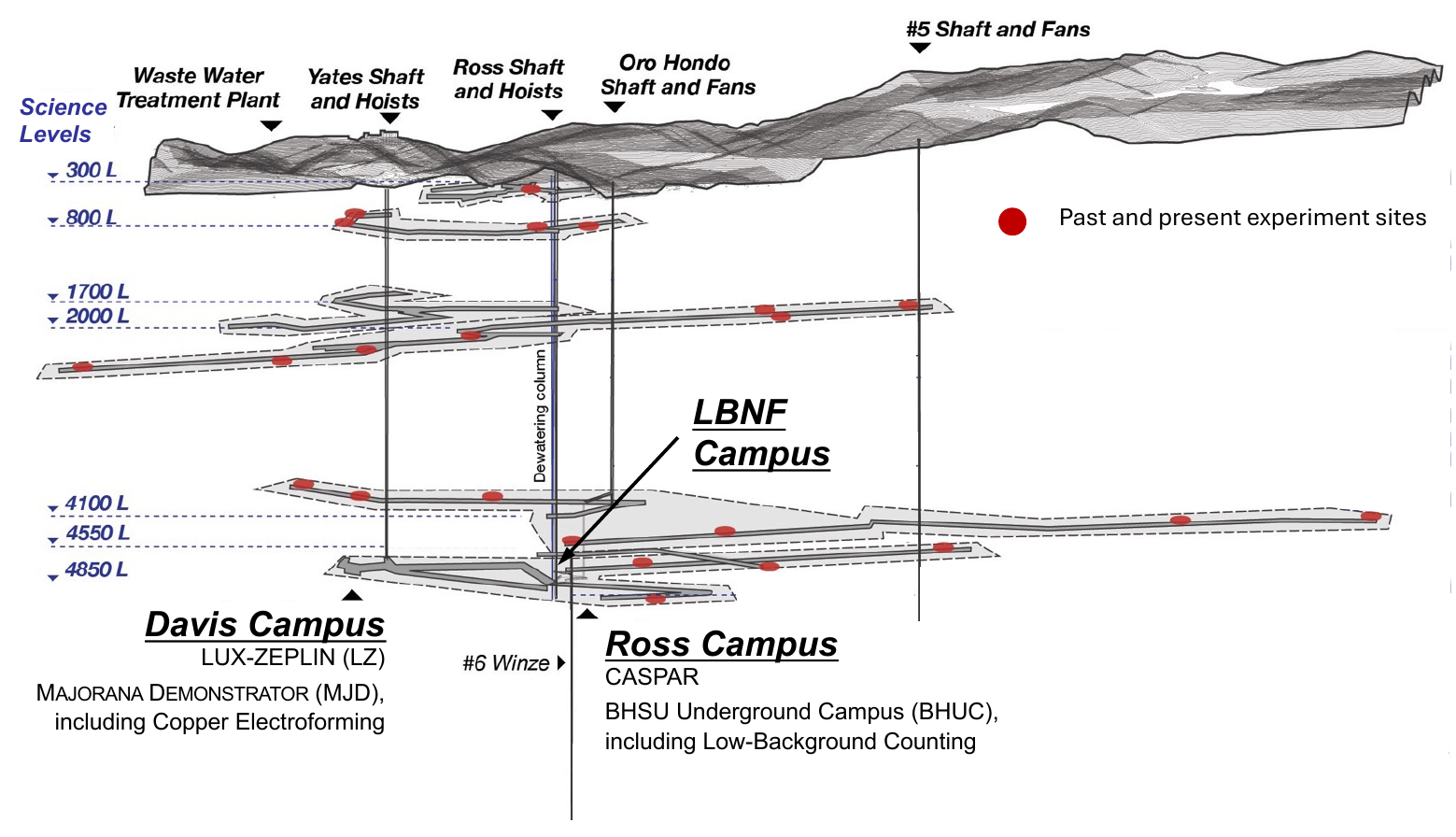}
    \caption{Simplified view of SURF's geography, including the main underground elevations for science and the two operating 4850L campuses; the location of the 4850L LBNF Campus is also indicated.}
    \label{fig:geography}
\end{figure}

Extending from the surface to more than 2450~meters below ground, approximately 600~km of tunnels and shafts were created during Homestake's 126~years of mining operations. The facility property comprises approximately 1~km$^{2}$ on the surface and more than 31~km$^{2}$ underground (a portion of which is represented in Figure~\ref{fig:geography}). Two main shafts -- the Ross and the Yates -- provide redundancy in terms of safe access and some services such as power and network communication. A series of pumping stations at various elevations in the Ross Shaft is used to pump ground water to the surface~\footnote{The average ground water inflow into the underground workings is 2760~lpm. After pumping ceased in June 2003~\cite{Murdoch-2012, Zhan_Duex-2010}, the mine filled with water until a high-water mark of 1381~meters below surface was reached in August 2008. Sustained pumping resumed in June 2008, and by May 2009 the water level dropped below the 4850L, which had been flooded for an estimated 16~months.}. Underground ventilation is provided primarily by the Oro Hondo fan and supplemented by the fan at \#5 Shaft, which currently draw a combined $\sim$990,000~m$^{3}$/h of fresh air underground via the Ross and Yates Shafts (sufficient for proposed future experiments); improvements are being evaluated for \#5 Shaft that could double the overall airflow.

In order to provide increased capacity to support the construction and operation of future experiments, state, private and federal funds were used to perform extensive renovations in the Ross Shaft. Started in August 2012, refurbishment of the Ross Shaft and hoists is complete and being leveraged for current experiment construction as well as potential future expansion. A similar refurbishment project is envisioned for the Yates Shaft.

Research activities at SURF are supported by facilities both on the surface as well as underground.

\subsection{Science Facilities -- Surface}
\label{sec:surface}

On the surface, the principal facility that directly serves science needs is the Surface Laboratory, which was first renovated starting in 2009 and provides approximately 210~m$^{2}$ of lab space (265~m$^{2}$ total). After a second renovation that was completed in July 2017, the Surface Laboratory facility includes two cleanrooms (total of more than 90~m$^{2}$), one of which is served by a commercial radon-reduction system capable of delivering 300~m$^{3}$/h and a measured reduction of 2200$\times$ at the output (770$\times$ inside the cleanroom). A new surface maintenance and support facility opened in 2021 that replaces the shipping and receiving warehouse located at the Ross Complex, consolidates maintenance capabilities and resources, provides office space as well as offering some staging and laydown space for research groups (the new facility was funded by a \$6.5M state investment and has a total footprint of 2415~m$^{2}$). A Waste Water Treatment Plant is used to treat water pumped from the underground (filtered material such as iron has been used in some SURF research efforts), and some experiment chemical processes are also performed there. 

SURF is the steward of an extensive drill core repository inherited from Homestake-Barrick that comprises a total of 39,760 boxes of core representing 2,688 drill holes (91~km total length). A database for the collection exists containing 58,000+ entries representing 1,740 drill holes. Homestake core holes extend to 3290~meters below surface where temperatures reach $\sim$75$^{\circ}$C~\cite{Mitchell-2023}. Additional core has been added over the years, including from DUSEL geotechnical investigations.

\subsection{Science Facilities -- Underground}
\label{sec:facilities_ug}

SURF offers access to a rock mass volume of 35~km$^{3}$, and with approximately 34~km maintained for access on 29 underground levels between the surface and the 5000-foot level (representing an area of 119,740~m$^{2}$), provides the largest geography for scientific purposes of any global underground laboratory. Space on seven primary levels identified for science activities as summarized in Table~\ref{tab:science_levels}.

\begin{table}[htbp]
\caption{Summary of key features for the primary SURF underground science levels, including information from a geological model~\cite{RoggenthenHart-2014} (meters of water equivalent = m.w.e.\@). Depth is represented by averaging values near the main access shafts.}
\label{tab:science_levels}
\begin{center}
\begin{tabular}{@{}r|r|r|r|l@{}}
\hline
{\bf Science}  & \multicolumn{2}{c|}{\bf Vertical Depth} & {\bf Accessible Footprint} & {\bf Services} \\
{\bf Level}    & {\bf (m)} & {\bf (m.w.e.\@)} & {\bf (Linear distance, m)} & \\
\hline
300L           &  130  &  350    &  940                   &                  \\ 
800L           &  270  &  750    & 2210                   &                  \\ 
1700L          &  550  & 1530    & 3040                   & Power, network   \\ 
2000L          &  650  & 1710    & 3050                   & in limited areas \\ 
4100L          & 1280  & 3600    & 2470                   &                  \\ 
4550L          & 1430  & 3970    & 1210                   &                  \\ 
4850L          & 1510  & 4300    & 5860, incl LBNF        & Significant services in labs, power     \\ 
               &       &         & (+ 420 Ph.B Expansion) & and network in other areas              \\
{\it 7400L}    & {\it 2260} & {\it 6460} & {\it TBD (nominal = 1490)} & {\it TBD (possible future)} \\
\hline
\end{tabular}
\end{center}
\end{table}

\begin{figure}[!htbp]
    \centering
    \includegraphics[width=0.9\textwidth]{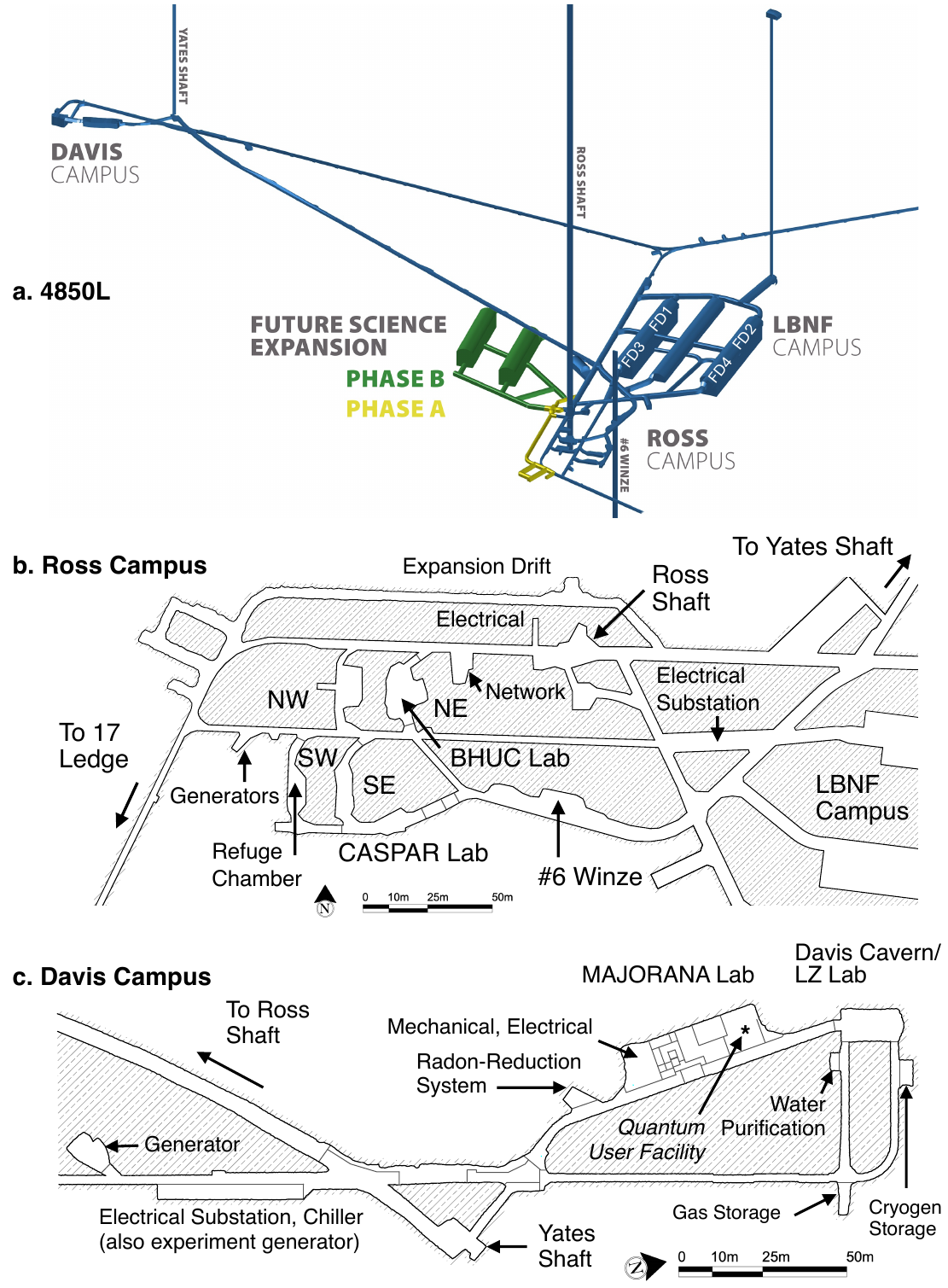}
    \caption{(\textit{a}) SURF current and proposed underground laboratory space on the 4850-foot level. Two new caverns in green are shown (100~m L $\times$ 20~m W $\times$ 24~m H) with a rock overburden similar to existing laboratories (1500~m, 4200~m.w.e.\@); Phase A Expansion excavation is complete. (\textit{b}) Ross Campus: four existing excavations are labeled NW, NE, SE, SW. The two western shop areas are currently in use as laboratories, whereas the two eastern shops are presently used for non-scientific purposes. (\textit{c}) Davis Campus: the main laboratory spaces and supporting areas are shown, including the proximity to the Yates Shaft. An asterisk indicates the location for the proposed Quantum User Facility (see Section~\ref{sec:quantum_user_facility}).}
    \label{fig:4850L}
\end{figure}

While significant geology and engineering spaces have been developed on the 1700L and 4100L~\footnote{Approximately 470~m$^{2}$ near the 4100L Yates Shaft has been outfitted to support geothermal and other engineering activities, including several drill holes. By the end of 2025, Caterpillar Inc.\ vacated space on the 1700L that is roughly equivalent to the current 4850L Ross Campus laboratory footprint, with heights near 6~m in some areas.}, the 4850L is the primary level for science at SURF and offers some of the deepest underground laboratory space in the world (see Figure~\ref{fig:LabVolumes}). Current experiments and infrastructure improvements are concentrated near the Yates and Ross Shafts where two well-furnished underground research campuses are located and where new excavation has been completed. Laboratory layouts are shown in Figure~\ref{fig:4850L}, where cleanroom spaces operate as low as class 10--100 with appropriate protocols; campus footprints (current and proposed) are summarized in Table~\ref{tab:footprint}. The main Refuge Chamber at the Ross Campus currently supports a maximum occupancy of 250 people, increased in May 2024 to support the projected construction peak in 2028/2029. Additional refuge provisions are also available at the Davis Campus to support 39 people. 

\begin{table}[htbp]
\caption{Area and volume footprints for various SURF underground laboratory spaces (including potential use of 1700L space). Total space includes access tunnels and facility support areas in addition to science laboratory areas. Quantities for proposed spaces are indicated in italics with up to two new 4850L laboratory caverns (see Figure~\ref{fig:4850L}) and the possibility of future development on the 7400L.}
\label{tab:footprint}
\begin{center}
\begin{tabular}{lrrrr}
\hline
{\bf Laboratory}                  & \multicolumn{2}{c}{\bf Science}   & \multicolumn{2}{c}{\bf Total}     \\
                                  & {\bf Area}      & {\bf Volume}    & {\bf Area}      & {\bf Volume}    \\
                                  & {\bf (m$^{2}$)} & {\bf (m$^{3}$)} & {\bf (m$^{2}$)} & {\bf (m$^{3}$)} \\
\hline
1700L CAT Campus                  &        923 &        3130  &            &               \\[0.2cm]
4850L Davis Campus        &       1018 &        4633  &      3017  &      11,354   \\
4850L Ross Campus                 &        920 &        3144  &      2653  &        8805   \\[0.2cm]  
4850L LBNF/DUNE (nominal)         &       9445 &     191,863  &    17,251  &     242,190   \\
4850L LBNF/DUNE (laser scan)      &       9958 &     198,871  &    19,288  &     248,217   \\[0.2cm]
{\it 4850L New 1$\times$100m Lab} & {\it 2011} & {\it 47,304} & {\it 3686} & {\it 73,293}  \\
{\it 4850L New 2$\times$100m Lab} & {\it 4022} & {\it 94,607} & {\it 6391} & {\it 129,274} \\[0.2cm]
{\it 7400L New 1$\times$75m Lab}  & {\it 3053} & {\it 28,152} & {\it 6733} & {\it 47,910}  \\
{\it 7400L New 2$\times$75m Lab}  & {\it $\sim$4178} & {\it $\sim$42,440} & {\it TBD} & {\it TBD} \\
\hline
\end{tabular}
\end{center}
\end{table}


\subsubsection{4850L Ross Campus}

The 4850L Ross Campus encompasses a set of four existing excavations near the Ross Shaft that were used as maintenance shops during mining activities, with a total footprint of 2653~m$^{2}$. Use of these former shops afforded an economical means to implement experiments in a timely manner --- a relatively modest \$4M of South Dakota funding was used to support the development of laboratory space. The layout shown in Figure~\ref{fig:4850L} illustrates various features of the Ross Campus, including the location of SURF's main Refuge Chamber, with backup power provided by a diesel generator.

The first formal underground laboratory developed as part of SURF's operation involved the Ross Campus NW area, which was renovated starting in December 2009 and operating as a copper electroforming laboratory 2011--2017, before being decommissioned in early 2018. Available following renovations that completed in August/September 2015, two spaces are currently configured as laboratories comprising 920~m$^{2}$, as indicated in Table~\ref{tab:footprint} (all four areas represent a combined footprint of 1407~m$^{2}$). Black Hills State University developed an underground campus in the NE space of the Ross Campus, including a cleanroom for hosting multidisciplinary research activities. A special epoxy paint was applied to surfaces in this space to reduce radon emanation from the cavern rock as well as from the concrete floor (however, the space is not currently operated with a low-radon environment since the incoming air has no radon abatement). A portion of a tunnel and a former maintenance shop were combined to create laboratory space in the SE area for the CASPAR experiment, where roughly 40\% of the space is used for Ross Campus utilities (chiller and some HVAC equipment). The rock hosting the Ross Campus consists mostly of the Poorman Formation (with a small portion of the Homestake Formation in the two north spaces).


\subsubsection{4850L Davis Campus}

A state-of-the-art laboratory complex called the Davis Campus has been constructed at the 4850L. It is near the Yates Shaft as shown in Figure~\ref{fig:4850L}, expanding the area surrounding the original Ray Davis cavern to a total footprint of 3017~m$^{2}$ (see Table~\ref{tab:footprint}), including a stainless steel tank that can be used for shielding (7.6~m diameter, 6.4~m high), a commercial water purification system (37.8~lpm) and a vacuum-swing-adsorption radon-mitigation system (nominal flow 150~m$^{3}$/h, $\sim$700$\times$ Rn reduction at the output)~\cite{Street:2015nwa}. Recently, a gaseous nitrogen generator was installed and commissioned at the Davis Campus, typically providing 60~slpm, replacing liquid nitrogen available from surface storage that can be used for boiloff nitrogen (with lower radon concentrations than boiloff). Two diesel generators serve the Davis Campus -- one for facility fire and life-safety systems and another dedicated for emergency xenon recovery. The Davis Campus represents a \$16M South Dakota commitment using state and private funds, supporting excavation that took place September 2009 -- January 2011 to increase laboratory space beyond the original cavern used for the Ray Davis Homestake Chlorine experiment and provide additional egress. Laboratory outfitting began in June 2011 and was complete in May 2012. The rock surrounding the Davis Campus is mostly Yates Amphibolite containing very low (sub-ppm) levels of U/Th; higher-activity rhyolite intrusives are also present in some areas. Based on experiment requirements, low-activity construction materials (including concrete and shotcrete for laboratory areas) and additional shielding (steel plating below the Davis Cavern tank) were selected for the Davis Campus; for details, see~\cite{Bkgd-Gamma_LZ, Heise:2015vza}.


\subsubsection{4850L LBNF Campus}

The DOE leased space both on the surface as well as underground in order to construct a new Long-Baseline Neutrino Facility (LBNF) at SURF to support and house the Deep Underground Neutrino Experiment (DUNE). Following a groundbreaking ceremony in July 2017, underground construction began at SURF in 2019 with the main excavation commencing mid-2021 and lasting 2.5~years. At the completion of the excavation phase in early 2024, a total of $\sim$720,000~tonnes of rock had been removed via the upgraded Ross hoists and deposited in the Open Cut, a large pit that was previously mined by Barrick/Homestake, creating a new 4850L LBNF Campus with a total footprint of 19,288~m$^{2}$ (see Table~\ref{tab:footprint}). Installation of utilities is currently underway. The design for the underground laboratory envisions four detector chambers (each 20~m wide $\times$ 28~m tall $\times$ 70~m long), each able to accommodate a 17.5-kt (10-kt fiducial volume) liquid argon detector. LBNF caverns are located in the Poorman geologic formation.
\begin{marginnote}[]
\entry{LBNF}{Long-Baseline Neutrino Facility}
\entry{DUNE}{Deep Underground Neutrino Experiment}
\end{marginnote}

\subsection{Characterization}

The SURF rock mass has been well characterized over the past 10+ years, including recent LBNF/DUNE studies as well as geology and geothermal research efforts, not to mention leveraging data from the Homestake Mining Company and associated historical studies. A geologic model has been constructed to incorporate the complex surface topology and various geologic formations as well as other features that characterize the underground environment~\cite{RoggenthenHart-2014}. Properties of the 12 relevant rock formations are summarized in Table~\ref{tab:geology}, and site-specific details such as geography coordinates, depth and densities are summarized in Table~\ref{tab:location}.

\begin{sidewaystable}[htbp]
\caption{Geological characteristics for 60-degree cones above various 4850L laboratories (Amph.\ = Amphibolite). References are listed for elemental abundances used in calculating the average number of protons, $<Z>$, and the average atomic mass, $<A>$, relevant for muon energy loss interaction cross sections; densities are from~\cite{RoggenthenHart-2014}, (W.\ Roggenthen, private communication, 2015).}
\label{tab:geology}
\centering
\begin{tabular}{lccccrrrrrrrr} 
\hline
{\bf Formation} & {\bf Density} & {\bf $<Z>$} & {\bf $<A>$} & {\bf Geochemical} & \multicolumn{8}{c}{\bf Percent of Cone Volume} \\
                &               &             &             & {\bf Reference}   & \multicolumn{2}{c}{\bf Davis Campus} & \multicolumn{2}{c}{\bf Ross Campus} & \multicolumn{4}{c}{\bf LBNF Campus} \\
                & {\bf (g/cm$^3$)}   & & & & {\bf LZ} & {\bf MJD} & {\bf CASPAR}  & {\bf BHUC} & {\bf FD1} & {\bf FD2} & {\bf FD3} & {\bf FD4} \\ 
\hline
Deadwood             & 2.43 &      &      &                                   &  0.1 &  0.1 &  0.3 &  0.3 &  0.2 &  0.2 &  0.3 &  0.3 \\
Grizzly              & 2.78 &      &      &                                   & 17.9 & 17.8 &  9.4 &  9.7 & 13.6 & 15.0 & 13.7 & 15.2 \\
Flagrock             & 2.98 &      &      &                                   &  9.6 &  9.7 &  9.1 &  9.1 & 10.4 & 10.8 & 10.7 & 11.1 \\
Flagrock Amph.\      & 2.98 &      &      &                                   &  4.9 &  4.8 &  3.0 &  3.1 &  3.3 &  3.4 &  3.1 &  3.2 \\
Northwestern         & 2.84 & 11.6 & 22.5 & \cite{10.1130/G50542.1}           & 12.1 & 12.0 &  7.7 &  7.9 &  8.7 &  8.9 &  8.3 &  8.5 \\
Northwestern Amph.\  & 2.98 &      &      &                                   &    0 &    0 &  0.0 &  0.0 &  0.0 &  0.0 &  0.0 &  0.0 \\
Ellison              & 2.73 & 11.7 & 23.5 & \cite{Caddey-1991, 10.5382/GB.07} & 23.8 & 24.2 & 40.5 & 39.7 & 37.0 & 36.0 & 38.5 & 37.5 \\
Homestake            & 3.26 & 13.0 & 26.7 & \cite{Caddey-1991, 10.5382/GB.07} &  4.2 &  4.3 &  6.9 &  6.8 &  5.7 &  5.4 &  5.8 &  5.4 \\
Poorman              & 2.86 & 12.0 & 24.2 & \cite{Caddey-1991, 10.5382/GB.07} & 17.6 & 17.6 & 16.9 & 17.3 & 14.6 & 13.7 & 13.2 & 12.4 \\
Yates Amph.\         & 2.93 & 12.5 & 25.1 & \cite{Caddey-1991}                &  3.4 &  3.2 &  0.1 &  0.1 &  0.2 &  0.2 &  0.1 &  0.1 \\
Phonolite (Tertiary) & 2.54 & 11.2 & 21.1 & \cite{Halvorson1980}              &  0.1 &  0.1 &  0.1 &  0.1 &  0.1 &  0.1 &  0.1 &  0.1 \\
Rhyolite (Tertiary)  & 2.54 & 11.2 & 21.1 & Assume same                       &  6.3 &  6.2 &  6.0 &  6.0 &  6.1 &  6.3 &  6.2 &  6.3 \\[-0.1cm]
                     &      &      &      & as above                          &      &      &      &      &      &      &      &      \\
\hline
\end{tabular}
\end{sidewaystable}

\begin{sidewaystable}[htbp]
\caption{Location details for specific SURF underground laboratories. Elevation values above sea level are referenced to the finished floor in each laboratory. Overburden rock density estimates at different cone angles for various underground locations at SURF, using a 3-dimensional geological model~\cite{RoggenthenHart-2014}. Angles are relative to the vertical above each specific site. These values supersede those in~\cite{Heise:2017rpu}.}
\label{tab:location}
\centering
\begin{tabular}{lcccccccccc} 
\hline
{\bf Location} & \multicolumn{3}{c}{\bf Geographic Coordinates}      & \multicolumn{2}{c}{\bf Vertical}   & \multicolumn{5}{c}{\bf Overburden Density} \\
               &  {\bf Latitude} & {\bf Longitude} & {\bf Elevation} & \multicolumn{2}{c}{\bf Rock Depth} & {\bf 0 deg} & {\bf 15 deg} & {\bf 30 deg} & {\bf 45 deg} & {\bf 60 deg} \\
               & {\bf (deg)} & {\bf (deg)} & {\bf (m)} & {\bf (m)} & {\bf (m.w.e)} & \multicolumn{5}{c}{\bf (g/cm$^3$)} \\ 
\hline
\multicolumn{10}{l}{\bf 1700L CAT Campus} \\
\hline
LHD Shop       & 103.7669444 W & 44.3560675 N & 1075 & 360 & 1070 & 2.931 & 2.877  & 2.831  & 2.814  & 2.787 \\ %
Motor Barn     & 103.7659810 W & 44.3558052 N & 1075 & 300 &  860 & 2.898 & 2.950  & 2.856  & 2.813  & 2.778 \\ %
\hline
\multicolumn{10}{l}{\bf 4850L Davis Campus} \\
\hline
LUX/LZ Detector & 103.7511871 W & 44.3533012 N & 108 & 1470 & 4210 & 2.870 & 2.866  & 2.848  & 2.833  & 2.828 \\ %
MJD Detector    & 103.7512998 W & 44.3529679 N & 115 & 1480 & 4260 & 2.882 & 2.867  & 2.848  & 2.832  & 2.828 \\ %
\hline
\multicolumn{10}{l}{\bf 4850L Ross Campus} \\
\hline
BHUC (NE)       & 103.7584188 W & 44.3462962 N & 115 & 1500 & 4320  & 2.871 & 2.752 & 2.772 & 2.803 & 2.821 \\ 
CASPAR (SE)     & 103.7587231 W & 44.3459461 N & 115 & 1500 & 4170  & 2.782 & 2.749 & 2.770 & 2.801 & 2.820 \\ 
\hline
\multicolumn{10}{l}{\bf 4850L LBNF Campus} \\
\hline
FD1 (NE)        & 103.7546341 W & 44.3459914 N & 97 & 1420 & 3990 & 2.801 & 2.768 & 2.798 & 2.804 & 2.817 \\ 
FD2 (SE)        & 103.7548042 W & 44.3450261 N & 97 & 1390 & 3870 & 2.788 & 2.768 & 2.796 & 2.802 & 2.815 \\ 
FD3 (NW)        & 103.7556732 W & 44.3460857 N & 97 & 1410 & 3940 & 2.823 & 2.757 & 2.790 & 2.803 & 2.818 \\ 
FD4 (SW)        & 103.7558433 W & 44.3451204 N & 97 & 1400 & 3890 & 2.767 & 2.759 & 2.788 & 2.801 & 2.816 \\ 
\hline
\multicolumn{10}{l}{\bf 4850L Expansion} \\
\hline
North Cavern    & 103.7562481 W & 44.3475724 N & 115 & 1500 & 4140 & 2.754 & 2.756 & 2.787 & 2.806 & 2.822 \\
South Cavern    & 103.7566145 W & 44.3470327 N & 115 & 1500 & 4170 & 2.775 & 2.754 & 2.783 & 2.804 & 2.821 \\ 
\hline
\end{tabular}
\end{sidewaystable}


SURF and other groups have collected data characterizing the facility in terms of various radioactive backgrounds, including assays of relevant rock formations: Yates Amphibolite rock, which is relatively low in radioactivity (0.22~ppm U, 0.33~ppm Th and 0.96\% K), Poorman rock formation, which is higher in natural radioactivity (2.58~ppm U, 10.48~ppm Th and 2.12\% K), and rhyolite intrusives, which have the highest activity of all SURF rocks evaluated (8.75~ppm U, 10.86~ppm Th and 4.17\% K) (Y-D.\ Chan, private communication, 2012; W.\ Roggenthen and A.R.\ Smith, private communication, 2008; A.R.\ Smith, private communication, 2007 and updates through 2014).  

Long-term underground ambient air radon data have been collected at various locations, and recent 12-month averages indicate 340~Bq/m$^{3}$ (Davis Campus) and 200~Bq/m$^{3}$ (Ross Campus), with a low baseline of 150~Bq/m$^{3}$. Excursions have been observed at both campuses, typically correlated with maintenance and ventilation changes. Plans to further control air flow in the Yates Shaft are expected to improve the Davis Campus radon values.

Other efforts to characterize physics backgrounds in a number of underground areas were carried out by various research groups: muons (800L, 2000L~\cite{Bkgd-Muon}, 4850L Davis Campus: (5.31\thinspace$\pm$\thinspace0.17) $\times$ 10$^{-5}$~muons~m$^{-2}$s$^{-1}$~\cite{Bkgd-Muon_MJD} and more recently (5.09\thinspace$\pm$\thinspace0.08$_{\rm stat}$\thinspace$\pm$\thinspace0.10$_{\rm sys}$)$\times$ 10$^{-5}$~muons~m$^{-2}$s$^{-1}$~\cite{LZ:2026kit} in the Davis Cavern; also~\cite{Thesis-Ihm2018} that interprets historical values~\cite{Cherry:1983dp} as noted in~\cite{Woodley:2024eln}), thermal neutrons (4850L Davis Campus: (1.7\thinspace$\pm$\thinspace0.1) $\times$ 10$^{-2}$~neutrons~m$^{-2}$s$^{-1}$ \cite{Bkgd-Neutron_Best}) and gamma rays (various~\cite{Bkgd-Gamma}, 4850L Davis Campus: (1.9\thinspace$\pm$\thinspace0.4)~gammas~cm$^{-2}$s$^{-1}$~\cite{Bkgd-Gamma_LZ}).

\subsection{Services}

SURF provides a number of important services that ultimately serve the wider underground science community.

\subsubsection{Low-Background Assays and Characterization}
\label{sec:low-bkgd}

SURF has established a National Laboratory-equivalent capacity to perform low-background assays in support of the U.S.\ and the global underground physics community. To that end, a consortium of institutions with various assay capabilities has formed. Successive generations of rare-process physics experiments require ever-lower material backgrounds to improve the sensitivity reach of detector technology.

Currently~\footnote{Low-background material assays began at SURF in 2013 with a USD instrument associated with the Center for Ultra-low Background Experiments at DUSEL (CUBED), a South Dakota 2010 Research Center~\cite{Keller:2010fye, Szczerbinska:2010zz}.} operating at the 4850L Ross Campus are six high-purity germanium gamma-ray low-background counters provided by three BHUC user groups (LBNL, LLNL, and University of Kentucky/University of Alabama) and managed through the Black Hills State University (BHSU) underground campus (BHUC)~\cite{Tiedt:2023gld, Mount:2017iam}. This suite of instruments, including two dual-crystal systems, is used to characterize the intrinsic radiopurity levels of materials used in the constructions of detectors or the facilities that host them. Uranium and thorium sensitivities on the order of 0.1~$\mu$Bq/kg ($\sim$10s~ppt) are typical (see Table~\ref{tab:lbc_sensitivities}). The performance of one detector (Ge-IV) is currently being evaluated at the Davis Campus, and installation at the Ross Campus is expected in 2026. The SOLO counter that operated at the Ross Campus was relocated to surface facilities at BHSU and is no longer operational. Since 2016, a total of more than 374 samples have been processed. In addition to past and current SURF physics experiments (LUX, LZ, MJD), many groups have benefited from the facility, including COHERENT~\cite{COHERENT:2023sol}, CUPID~\cite{TROTTA2024169657}, DAMIC-M~\cite{DAMIC-M:2024ooa}, DUNE/ProtoDUNE~\cite{ManzanillasVelez:2024bms}, IceCube~\cite{IceCube:2022mng}, LEGEND~\cite{Zuzel:2025lsr}, nEXO~\cite{Adhikari_2022}, NEXT~\cite{NEXT:2025yqw}, TESSERACT~\cite{TESSERACT:2025tfw}, Theia~\cite{Theia-Snowmass} and XLZD~\cite{XLZD:2024nsu}. Local universities have some additional material screening capabilities: ICP-MS (BHSU) and radon-emanation characterization and alpha counting (SD Mines). In the coming years, additional high-purity germanium detectors are being considered for the BHUC facility.
\begin{marginnote}[]
\entry{BHUC}{Black Hills State University Underground Campus}
\end{marginnote}

\begin{table}[htbp]
\caption{ Low-background counter sensitivities~\cite{Tiedt:2023gld, Mount:2017iam}; see also~\cite{LZ:2020fty}.}
\label{tab:lbc_sensitivities}
\begin{center}
\begin{tabular}{lcccl}
\hline
{\bf Detector} & {\bf Ge}      & {\bf [U]}    & {\bf [Th]}    & {\bf Status/} \\
{\bf (Group)}  & {\bf Crystal} & {\bf mBq/kg} & {\bf mBq/kg}  & {\bf Comment} \\
\hline
Maeve   & 2.2 kg, & 0.1   & 0.1    & Production assays.\\
(LBNL)  & p-type  & (10~ppt) & (25~ppt) & Relocated from \\
        & ($\epsilon$=85\%) & & & Oroville, old Pb \\
        & & & & inner shield. \\[0.1cm]

Merlin  & 2.2 kg, & 0.2   & 0.2    & Commissioning. \\
(LBNL)  & n-type  & ($\sim$20~ppt) & ($\sim$50~ppt) & \\
        & ($\epsilon$=115\%) & & & \\[0.1cm]

Morgan  & 2.1 kg, & 0.2   & 0.2    &  Production assays. \\
(LBNL)  & p-type  & (20~ppt) & (50~ppt) & \\
        & ($\epsilon$=85\%) & & & \\[0.1cm]

Mordred & 1.3 kg, & 0.7 & 0.7  & Production assays. \\
(USD/   & n-type  & (60~ppt) & (175~ppt) & Shield access \\
CUBED,  & ($\epsilon$=60\%) &&& upgrade. \\
LBNL)   &&&& \\[0.1cm]

Dual HPGe    & 2$\times$2.1 kg, & $\sim$0.01 & $\sim$0.01 & Operating.  \\
``Twins''    & p-type & ($\sim$1~ppt) & ($\sim$1~ppt) & Flexible shield \\
(LBNL,BHSU,  & ($\epsilon$=2$\times$120\%) &&& configuration. \\
UCSB)        &&&& \\[0.1cm]

Ge-IV      & 2.0 kg, & 0.04     & 0.03    & Installation \\
(Alabama,  & p-type  & (3~ppt)  & (8~ppt) & underway. \\
Kentucky)  & ($\epsilon$=111\%) &&& Vertical design \\
           &&&& w/ gantry and hoist. \\[0.1cm]

Dual HPGe  & 2$\times$1.1 kg, & $<$0.1 & $<$0.1 & Production assays. \\
``RHYM+    & p-type & ($<$10~ppt) & ($<$25~ppt) & BEGe low-E $^{210}$Pb \\
RESN''     & ($\epsilon$=2$\times$65\%) &&& ($<$2 mBq/kg). \\
(LLNL)     &&&& \\
\hline
\end{tabular}
\end{center}
\end{table}

\subsubsection{Electroforming}
\label{sec:e-forming}

Production of electroformed copper is also performed at the facility (average U, Th decay chain $\leq$~0.1~$\mu$Bq/kg). The {\sc Majorana} collaboration has produced electroformed copper at SURF since mid-2011, where a total of $\sim$2500~kg of electroformed copper was produced for the {\sc Majorana Demonstrator}~\cite{Abgrall:2025tsj} during a period of approximately 4 years. Four baths currently operate at the Davis Campus, and the installation of two additional baths is underway, with the possibility to expand to eight baths being evaluated. Ongoing copper electroforming is nominally for LEGEND use, but materials have been provided to other projects. Electroforming activities are expected to continue at SURF into at least the early 2030s.

\subsubsection{Xenon}

SURF manages 1.5M~liters of xenon purchased through state foundation investments, further purified by the LZ collaboration to remove Kr to $\sim$100s~ppq levels.


\section{SCIENCE PROGRAM}

Building on the legacy of the Ray Davis chlorine solar-neutrino experiment that was conducted at the Homestake Mine over several decades~\cite{Cleveland:1998nv}, SURF's current science program owes much to community planning efforts that started in 2000~\cite{Mitchell-2023} and the first formal call for letters of interest in 2005. Science efforts started at SURF in 2007 during re-entry into the facility and have grown steadily over the past 18 years. To date, 71 groups have conducted research on various levels, ranging from the surface to the 5000L, where several specific SURF attributes are attractive to researchers in various disciplines~\footnote{While this article is nominally a review of nuclear and particle science, the full breadth of science served by SURF is included for completeness and in the spirit of broad scientific connections.}:
\begin{itemize}
\item General: Significant footprint (see Table~\ref{tab:science_levels}; other sites may be accessed on a case-by-case basis). Services (power, network) to some locations on most levels.
\item Physics: SURF provides a low-background environment (reduced flux of cosmic-ray muons and low vibrational noise) to study rare processes. Average rock overburden of $\sim$4300 m.w.e.\ at existing laboratories on 4850L (other levels available). Hard rock with geotechnical properties conducive to constructing large caverns; also, host rock with relatively low uranium/thorium levels.
\item Biology: Isolation from surface, variety of locations that result in different environmental conditions (temperature, humidity), variety of niches (different materials and rock geochemistry, various trace gases, access to water from various sources including isolated ground waters as old as tens of thousands of years, etc.).
\item Geology: Variety of geologic environments / rock formations (permeability, porosity, chemistry) as well as a variety of rock conditions (stress, temperature, etc.). The SURF drill core repository is a unique resource (see Section~\ref{sec:surface}).
\item Engineering: A variety of environments for testing real-world applications (especially related to mining), stress at depth for some testing.
\end{itemize}

SURF's strategic direction has a ten-year horizon and is largely guided by U.S.\ physics community efforts such as the recent 2023 Particle Physics Prioritization Project Plan (P5)~\cite{P5:2023wyd}. The Nuclear Science Advisory Committee (NSAC) 2023 Long Range Plan~\cite{NSAC2023} also has strong support for current and future science opportunities at SURF. To address SURF's multi-disciplinary scientific landscape, a steering committee has been convened to develop a SURF Scientific Strategic Plan for Non-Physics, which is expected to complete its report in early 2026.
\begin{marginnote}[]
\entry{P5}{Particle Physics Prioritization Project Plan}
\entry{NSAC LRP}{Nuclear Science Advisory Committee Long Range Plan}
\end{marginnote}

A total of 30 research programs are ongoing, spanning the fields of physics, geology, biology and engineering as summarized in Table~\ref{tab:science_programs}. The current science program sees 450 researchers active onsite at SURF from a total pool of more than 2400 collaborators, representing efforts from 319 institutions in 55 countries. Since being established a dedicated science laboratory in 2007, approximately 850 researchers have been active at SURF.

\begin{table}[htbp]
\caption{Current SURF scientific research programs. Proprietary groups are indicated with ($\dagger$).}
\label{tab:science_programs}
\begin{center}
\begin{tabular}{llll}
\hline
\multicolumn{1}{c}{\bf Experiment} & \multicolumn{1}{l}{\bf Description} & \multicolumn{1}{l}{\bf Location} & \multicolumn{1}{l}{\bf References} \\
\hline
\multicolumn{4}{l}{\bf Physics} \\
\hline
LZ & Dark matter using Xe & 4850L & {\scriptsize\cite{Akerib:2025xla, LZ:2024zvo, LZ:2022lsv, LZ:2019sgr, LZ:2019qdm, LZ:2020fty}} \\
(also XLZD R\&D) &&&\\[0.08cm]
MJD & 0$\nu\beta\beta$ / $^{180m}$Ta using Ge & 4850L & 
{\scriptsize\cite{Majorana:2024asd, Majorana:2022mrm, Majorana:2022gtu, Majorana:2023ecz, Majorana:2022udl, Majorana:2022bse}} \\
(also LEGEND support) &&& \\[0.08cm]
CASPAR & Nucleosynthesis & 4850L & {\scriptsize\cite{Borgwardt:2023rqi, Frentz:2022ucg, Shahina:2022xwg, Dombos:2022bph, Olivas-Gomez:2022tro}} \\[0.08cm]
Bkgd Characterization & Neutron, gamma & 4850L & \\[0.08cm]
DUNE & Neutrino measurements & 4850L & {\scriptsize\cite{DUNE:2024ptd, DUNE:2022aul, DUNE:2020lwj}} \\[0.08cm]
BHUC & Low-background assays & 4850L & {\scriptsize\cite{Tiedt:2023gld, Mount:2017iam}} \\
(users below) &&& \\[0.08cm]
\multicolumn{1}{r}{Berkeley}         & 5$\times$ HPGe systems & 4850L & {\scriptsize\cite{Smith:2015aoa, Thomas:2015gla}} \\[0.08cm]
\multicolumn{1}{r}{Alabama/Kentucky} & 1$\times$ HPGe system  & 4850L & \\[0.08cm]
\multicolumn{1}{r}{LLNL}             & 1$\times$ HPGe system  & 4850L & \\[0.08cm]
\hline
\multicolumn{4}{l}{\bf Geology} \\
\hline
CUSSP & Geothermal energy & 4100L & {\scriptsize\cite{CUSSP-1, CUSSP-2}} \\[0.08cm]
DEMO-FTES & Geothermal energy & 4100L & {\scriptsize\cite{FTES2025, FTES2024-2, FTES2024-1}} \\[0.08cm]
3D DAS & Seismic monitoring & 4550L, 4850L & {\scriptsize\cite{3D_DAS:2025, 10.1785/0220230180, 3D_DAS:2023, 3D_DAS:2022}} \\[0.08cm]
Core Archive ($\dagger$) & Gold deposits & Surface & \\[0.08cm]
USGS Hydrogravity & Hydrology & Many levels & {\scriptsize\cite{Hydrogravity2015}} \\[0.08cm]
Transparent Earth & Seismic monitoring & Several levels & 
{\scriptsize\cite{10.56952/ARMA-2025-0730}} \\[0.08cm]
Black Hills Seismic & Seismic monitoring & 4100L & \\[0.08cm]
\hline
\multicolumn{4}{l}{\bf Biology} \\
\hline
DeMMO & Geomicrobiology & Surface, 800L & {\scriptsize\cite{10.3389/fmicb.2024.1455594, 10.1128/msystems.00966-23, DeMMO2023, 10.1128/AEM.00832-21, DeMMO2021-1},} \\
      &                 & 2000L, 4100L, 4850L & {\scriptsize\cite{DeMMO2021-2, Osburn2020.09.15.298141, DeMMO2020, DeMMO2019}}\\[0.08cm]
Biodiversity (BHSU) & Microbiology & Many levels & {\scriptsize\cite{Bergmann:2020, Bergmann:2018}}\\[0.08cm]
Biofuels (SDSMT) & Biofuels & Many levels & {\scriptsize\cite{microorganisms9010113, GOVIL2019105730, DHIMAN2018270, SHRESTHA2018318, doi:10.1128/genomea.00405-17}} \\[0.08cm]
2D Best & Biofilms & 4850L & \\[0.08cm]
m-sense & Microbial variations & Several levels & {\scriptsize\cite{DEVADIG2025112988, msense:2025}} \\[0.08cm]
Chemistry (BHSU) & Characterization & 4850L & {\scriptsize\cite{Chemistry:2017}} \\[0.08cm]
Delavie Sciences ($\dagger$) & Novel proteins & 4850L & \\[0.08cm]
DULIA-Bio/REPAIR & Yeast in low radiation & Surface, 4850L & \\[0.08cm]
\hline
\multicolumn{4}{l}{\bf Engineering} \\
\hline
Caterpillar ($\dagger$) & Mining equipment & 1550L, 1700L & \\[0.08cm] 
Thermal Breakout & Technology for stress & 4100L, 4850L & {\scriptsize\cite{10.2118/201195-PA}} \\[0.08cm]
AMD ($\dagger$) & Chip error testing & 4850L & \\[0.08cm]
Shotcrete & Mining safety & Surface & \\[0.08cm]
Environment Monitoring & Air-flow data & Many levels & \\[0.08cm]
MAP & Remediation w/ microbes & 1700L, 2000L & {\scriptsize\cite{Tukkaraja:2025}} \\[0.08cm]
\hline
\end{tabular}
\end{center}
\end{table}

\subsection{Dark Matter: LUX-ZEPLIN}

The LUX-ZEPLIN (LZ) experiment~\cite{LZ:2019sgr} is the upgraded successor to the successful Large Underground Xenon (LUX) project that operated at the 4850L Davis Campus from 2013--2016 using approximately 300~kg of xenon~\cite{LUX:2016ggv}. The LZ collaboration is performing a direct search for weakly interacting massive particles (WIMPs), the leading candidate for dark matter, using 10~tonnes of xenon (5.5-tonne fiducial mass) in dual-phase mode within an ultra-pure titanium cryostat immersed in a large tank containing ultra-pure water and a gadolinium-loaded liquid scintillator (linear alkyl-benzene) veto. The LZ detector has been operating underground at the SURF 4850L Davis Campus since December 2021.
\begin{marginnote}[]
\entry{LZ}{LUX-ZEPLIN}
\entry{LUX}{Large Underground Xenon}
\entry{ZEPLIN}{ZonEd Proportional scintillation in LIquid Noble gases}
\entry{WIMP}{Weakly Interacting Massive Particle}
\entry{XLZD}{XENON LZ DARWIN}
\end{marginnote}

The first WIMP-search results were announced for the LZ experiment in July 2022~\cite{LZ:2022lsv}, with updated results last announced in August 2024~\cite{LZ:2024zvo} (both world-leading). A new low-energy analysis was published that includes the experiment’s sensitivity to background $^{8}$B solar neutrinos via coherent elastic neutrino-nucleus scattering (CE$\nu$NS)~\cite{Akerib:2025xla}.
Currently, LZ aims to reach a projected sensitivity to spin-independent interactions of $\sigma_{LZ}$ = 3$\times$10$^{-48}$~cm$^{2}$ for a 40~GeV/c$^{2}$ mass WIMP based on operating the LZ detector for 1000 live days (until $\sim$2028). Due to the large xenon target, the LZ experiment also has sensitivity to the 0$\nu\beta\beta$ decay of $^{136}$Xe~\cite{LZ:2019qdm}. The majority of the experiment's xenon inventory is managed by the SDSTA through foundation investments ($\sim$80\% of the 1.9M-liter total) and is secured through September 2028. SDSTA's Xe inventory is expected to become available for long-term community use, and LZ is considering an extended program beyond its current planned operation, with concepts such as HydroX to increase sensitivity to low-mass dark matter and CrystaLiZe to isolate background sources~\cite{Gibbons:2025pza, HydroX:2025nxn} as well as possible neutrino studies with electron-capture sources.
\begin{marginnote}[]
\entry{CE$\nu$NS}{coherent elastic neutrino-nucleus scattering}
\end{marginnote}

Plans are underway for more sensitive searches to regions of WIMP parameter space where solar and atmospheric neutrinos constitute an appreciable background. Advocates for a next-generation liquid Xe observatory for dark matter and neutrino physics, combining U.S.\ and European dark matter researchers, formalized a new collaboration called XLZD in 2022~\cite{XLZD:2024pdv, Aalbers:2022dzr}, recognizing contributions from the XENON, LZ and DARWIN xenon-based experiment collaborations and proposing a target of up to 100~tonnes of Xe. Based on published requirements~\cite{XLZD:2024nsu}, five underground laboratories, including SURF, are contending to host this ambitious project starting in the early-2030s timeframe, aligned with SURF expansion plans. The P5 report has very strong support for Generation-3 dark matter experiments at SURF: ``Investment in the expansion of SURF, taking advantage of the DUNE excavation infrastructure and potential private funding, would enable such siting.''

\subsection{Neutrinos: {\sc Majorana Demonstrator}}

For more than a decade, the {\sc Majorana Demonstrator} (MJD) experiment has investigated rare decays at SURF. The main search was focused on neutrinoless double-beta decay (0$\nu\beta\beta$) and employed approximately 44~kg of germanium detectors (30~kg was enriched to 88\% $^{76}$Ge) in two cryostats protected inside a Cu/Pb/HDPE shielding enclosure ($\sim$66~tonnes) with an active muon veto. Production science data were collected from 2015--2021, reaching the exposure goal of 65~kg-years for a half-life limit of T$_{1/2} >$ 8.3$\times$10$^{25}$~years~\cite{Majorana:2022udl}; a search for transitions to excited states of $^{76}$Se was also performed~\cite{Majorana:2024asd}. All enriched Ge detectors were removed from the {\sc Demonstrator} in 2021 for use in the next-generation Ge-based program: the Large Enriched Germanium Experiment for Neutrinoless $\beta\beta$ Decay (LEGEND) experiment, with an initial phase using up to 200~kg Ge that is now taking data at LNGS~\cite{Saleh:2025bdy}, leading to a future 1000~kg version~\cite{Zuzel:2025lsr}.
\begin{marginnote}[]
\entry{MJD}{{\sc Majorana Demonstrator}}
\entry{0$\nu\beta\beta$}{neutrinoless double-beta decay}
\entry{LEGEND}{Large Enriched Germanium Experiment for Neutrinoless $\beta\beta$ Decay}
\end{marginnote}

Starting in May 2022 using the remaining natural Ge detectors and {\sc Demonstrator} shield, the collaboration set out to study the decay lifetime of $^{\rm 180m}$Ta using 17~kg of tantalum. Initial world-leading search results have been published~\cite{Majorana:2023ecz}. Over the lifetime of the MJD experiment, a number of other significant results were published, including searches for dark matter~\cite{Majorana:2022gtu, Majorana:2022bse} and other Beyond Standard Model physics~\cite{Majorana:2022mrm}. The collaboration completed operations with the {\sc Demonstrator} in Summer 2025, and decommissioning is expected to be complete in 2026. As noted in Section~\ref{sec:e-forming}, the {\sc Majorana} collaboration invested significant resources to produce ultra-pure underground electroformed copper, which continues at SURF in support of the LEGEND experiment as well as serving the needs of the larger scientific community.

The neutrinoless double-beta decay community is planning for at least one ton-scale instrument sensitive to half-lives greater than 10$^{28}$~years. To host a ton-scale 0$\nu\beta\beta$ experiment at SURF, additional space is required. While new excavation is planned at SURF (see Section~\ref{sec:future}), the associated timeline for development does not currently support projected schedules for experiment construction. In addition, the community is already planning for one more leap in sensitivity beyond the ton scale, and an initial concept for a 6000-kg version of LEGEND~\cite{LEGEND-6000} is being considered, including use of underground argon with reduced levels of cosmogenically-produced long-lived radioactive isotopes $^{39}$Ar and $^{42}$Ar~\cite{GlobalArgonDarkMatter:2024wtv}, with a projected half-life sensitivity of T$_{1/2}$ $\sim$10$^{29}$~years. A deep site like SURF ensures that cosmogenic activation backgrounds such as $^{\rm 77m}$Ge do not have a significant impact on the sensitivity of the experiment. In particular, the recent NSAC LRP report mentions that ``The premier underground laboratory in the United States is Sanford Underground Research Facility (SURF) in Lead, South Dakota.... As SURF continues to expand, it would be suited to host the next generation of neutrinoless double-beta decay experiment.''

\subsection{Nuclear Astrophysics: CASPAR}

The Compact Accelerator System for Performing Astrophysical Research (CASPAR) is one of three deep underground laboratories for nuclear physics in the world, studying stellar nuclear fusion reactions using a 1-MV Van de Graaff accelerator capable of producing high-intensity ($\sim$200~$\mu$A) proton and alpha beams with energies between 150~keV and 1.0~MeV, low-energy regions relevant to key astrophysical processes including hydrogen burning and stellar helium burning in environments ranging from novae to the first generation of stars formed after the Big Bang. Furthermore, neutron production reactions will help interpret various nucleosynthesis channels, which are essential to understanding the origin of many of the heavy elements in our Universe. The underground environment at SURF shields cosmic-ray muons that can themselves create backgrounds for CASPAR as well as induce reactions in the rock that can produce other backgrounds such as neutrons.
\begin{marginnote}[]
\entry{CASPAR}{Compact Accelerator System for Performing Astrophysical Research}
\end{marginnote}

Accelerator components were relocated from the University of Notre Dame in Summer 2015, and since February 2018 a number of data campaigns have been conducted at the 4850L Ross Campus laboratory, including the following targets: $^{7}$Li, $^{11}$B, $^{14}$N, $^{18}$O, $^{20}$Ne, $^{22}$Ne (gas, solid), $^{27}$Al~\cite{Borgwardt:2023rqi, Frentz:2022ucg, Shahina:2022xwg, Dombos:2022bph, Olivas-Gomez:2022tro}. In particular, the $^{14}$N(p,$\gamma$)$^{15}$O reaction is relevant for stellar CNO reactions. Measurement of the CNO neutrino flux offers an independent measure of the metallicity of the solar core~\cite{BOREXINO:2023ygs, BOREXINO:2020aww}, for which low-energy cross sections for CNO reactions are needed.

CASPAR experimental operations temporarily halted in March 2021 due to nearby LBNF construction, resuming for a second phase of operations in Summer 2025. Measurements planned for the next phase include nucleosynthesis reactions in first stars (using $^{6}$Li, $^{10}$B and $^{11}$B targets), especially relevant in light of interpretations of primordial lithium abundances in the early Universe by the James Webb Space Telescope~\cite{Carniani2024}. The first measurement in the second phase will be $^{19}$F(p,$\gamma$)$^{20}$Ne, which may provide a link to advanced hydrogen burning cycles that are able to produce elements up to calcium, and possibly explain the observation of a relatively large abundance of calcium in otherwise ultra-metal-poor (very old) stars. The experimental program associated with CASPAR Phase 2 is expected to last for approximately three years, until $\sim$2028.

The CASPAR collaboration is proposing to continue their nuclear astrophysics program in a third phase of operation, with a possible electrical upgrade further optimizing low-energy measurements that could begin in the early- to mid-2030s using a high voltage platform in the 200--300~kV range. In the NSAC LRP, the nuclear science community recognizes the value of a deep underground site for nuclear astrophysics research to address the broad range of experimental questions associated with the nucleosynthesis in stars: ``Answering the remaining questions about the nuclear processes in the Sun will require measurements of nuclear rates at low-energy accelerators..., including the deep underground Compact Accelerator for Performing Astrophysical Research (CASPAR).''

\subsection{Black Hills State University Underground Campus}

As noted in Section~\ref{sec:low-bkgd}, the Black Hills State University Underground Campus (BHUC) manages low-background counting efforts at SURF for the national and international community through a consortium of users. The Ross Campus facility provides potential space for modest-scale groups beyond low-background counting, and a portion of the laboratory structure is available for non-physics activities (such as initial observations of biology samples). One of the main goals of the BHUC is to engage undergraduate students in research.

\subsection{Neutrinos: DUNE}

LBNF's far-site location at SURF will house the DUNE experiment~\cite{DUNE:2023nqi, DUNE:2023jco, DUNE:2022aul, DUNE:2020lwj}, the largest international science project ever to be constructed on U.S.\ soil. With more than 1,500 scientists from 211 laboratories and universities in 40 countries (including CERN), DUNE will be a world-class neutrino observatory and nucleon decay detector, designed to answer fundamental questions about the nature of elementary particles and their role in the evolution of the universe. In particular, DUNE science will touch on such compelling unanswered puzzles as neutrino mass hierarchy and CP violation (matter/antimatter asymmetry), as well as many other topics, including detecting neutrinos from low-energy astrophysical sources such as a supernova explosion~\cite{DUNE:2024ptd}.

The DUNE experiment will consist of a far detector to be located on the 4850L of SURF that will observe neutrinos generated approximately 1300~km away at Fermilab using an upgraded accelerator beam (up to 2.4~MW), producing a broad neutrino energy spectrum. The detectors at SURF will be very large, modular liquid argon time-projection chambers with a total of 70~kt (40~kt fiducial) mass. Liquid argon technology will make it possible to reconstruct neutrino interactions with image-like precision and unprecedented resolution. 

DUNE science is anticipated to start in early 2030, with the FNAL beam upgrade complete in 2031 and the near detector currently expected to be online in 2032. The LBNF/DUNE project is being managed in two phases: Phase I consists of two far detectors (vertical drift and horizontal drift), and Phase II is likely to include at least one vertical drift detector~\cite{DUNE:2025gyl,DUNE:2024wvj}. The primary objective of the DUNE experiment is a set of precise measurements of the neutrino mixing matrix parameters, $\theta_{\rm 23}$, $\theta_{\rm 13}$, $\Delta$m$^{2}_{32}$, and $\delta_{\rm CP}$, to establish CP violation over a broad range of possible values of $\delta_{\rm CP}$, and to search for new physics in neutrino oscillations. DUNE Phase II, with its full 1000~kt$\cdot$MW$\cdot$yr exposure (achieved after 15 years of beam), will enable DUNE to establish CP violation at $>$3$\sigma$ over 75\% of possible $\delta_{\rm CP}$ values.

The 2023 P5 report reaffirms this vision and recognizes that with the Fermilab Accelerator Complex Evolution - Main Injector Ramp and Targets (ACE-MIRT) upgrade, three DUNE far detectors based on liquid argon technology will achieve the DUNE science goals. A fourth detector could be envisioned as a ``Module of Opportunity'', providing an expanded program with excellent prospects for future underground physics.
\begin{marginnote}[]
\entry{ACE-MIRT}{Accelerator Complex Evolution - Main Injector Ramp and Targets}
\end{marginnote}

\subsection{Geology \& Geomechanics: Enhanced Geothermal Systems, Seismic Studies}
\label{sec:geology}

Geological and geomechanical investigations directly bear on critical topics such as energy production and infrastructure stability that benefit from access to a variety of rock types, pressures, and complexly folded structures.

The Demonstration of Fracture Thermal Energy Storage (DEMO-FTES) experiment~\cite{FTES2025, FTES2024-2, FTES2024-1} studies enhanced geothermal system (EGS) and fracture thermal energy storage (FTES) effects on 10-meter scales. The DEMO-FTES experiment conducted water-flow tests starting in late 2024, leveraging considerable DOE infrastructure investments on the 4100L, including 11 drill holes (180--265~m long) and some existing instrumentation. DEMO-FTES efforts are expected to wrap up by the end of 2025.
\begin{marginnote}[]
\entry{EGS}{Enhanced Geothermal Systems}
\entry{FTES}{Fracture Thermal Energy Storage}
\entry{kISMET}{Permeability (k) and Induced Seismicity Management for Energy Technologies}
\entry{CUSSP}{Center for Understanding Subsurface Signals and Permeability}
\end{marginnote}

In conjunction with the DEMO-FTES project, a new DOE Energy Earthshot Research Center -- the Center for Understanding Subsurface Signals and Permeability (CUSSP) -- will use the 4100L site to focus on important aspects of EGS such as understanding and detecting the chemo-mechanical interactions that control long-term flow through fracture networks~\cite{CUSSP-1, CUSSP-2}. Initial CUSSP efforts are underway at SURF focused on further characterizing the 4100L testbed. The CUSSP experiment aims to collect measurements for several years, and a 2025 DOE call for proposals is expected to promote continued use of SURF geothermal facilities. Note that SURF can also provide field demonstration opportunities in support of DOE's flagship EGS effort called the Frontier Observatory for Research in Geothermal Energy (FORGE). 
\begin{marginnote}[]
\entry{FORGE}{Frontier Observatory for Research in Geothermal Energy}
\end{marginnote}

Current geothermal efforts build on results and infrastructure provided by earlier projects hosted at SURF over the time period 2016--2022. These predecessor experiments include EGS Collab~\cite{EGS-Yu2024, EGS-Overview2024, EGS-Overview2020} and Permeability (k) and Induced Seismicity Management for Energy Technologies (kISMET)~\cite{kISMET-Overview2017}. 

The 3D Distributed Acoustic Sensing (3D DAS) collaboration is performing enhanced seismic monitoring using backscattered laser pulses in optical fiber. Starting in 2022, the 3D DAS collaboration worked with SDSTA personnel to install a total of 3100~m of cable in a ramp system. Bragg grating optical fiber was installed in early 2025 for similar measurements in the new 4850L Expansion drift. An earlier Geoscience Optical Extensometers and Tiltmeters (GEOX$^{\rm TM}$) collaboration used optical fibers on the 4100L in 2011 to collect temperature and rock strain measurements, including effects due to active loading~\cite{GAGE2014350}.
\begin{marginnote}[]
\entry{3D DAS}{Three-dimensional Distributed Acoustic Sensing}
\entry{GEOX$^{\rm TM}$}{Geoscience Optical Extensometers and Tiltmeters}
\end{marginnote}

\subsection{Biology and Geobiology: Astrobiology and Extremophiles}

Important questions in life science, such as the conditions of life, the extent of life and ultimately the rules of life, are being addressed underground at SURF. Biology researchers take full advantage of SURF's large footprint by gathering samples from a number of underground levels and areas with different temperatures and geologic mineralogies. Various groups focus on the diversity of life, including understanding microbial metabolism in rock-hosted ecosystems, while others address engineering applications such as improvements to biofuel production and carbon sequestration. Many samples cannot be cultured in the laboratory, leading groups to perform {\it in situ} culturing of microbial life.

The Deep Mine Microbial Observatory (DeMMO) collaboration~\cite{10.3389/fmicb.2024.1455594, 10.1128/msystems.00966-23, DeMMO2023, 10.1128/AEM.00832-21, DeMMO2021-1, DeMMO2021-2, Osburn2020.09.15.298141, DeMMO2020, DeMMO2019} has outfitted six drill holes on four underground elevations for long-term water sampling and monitoring. Current activities resulted from a precursor project, Life Underground -- NASA Astrobiology Institute~\cite{10.3389/fenrg.2019.00121, 10.3389/fmicb.2018.01993, 10.1111/1758-2229.12563} that was funded for several years at SURF (2014--2018), including drill core analysis performed using Jet Propulsion Laboratory's {\it in situ} SHERLOC laser spectrometer~\cite{SHERLOC}, the technology concept used on the Mars Perseverance rover.
\begin{marginnote}[]
\entry{DeMMO}{Deep Mine Microbial Observatory}
\entry{SHERLOC}{Scanning Habitable Environments with Raman \& Luminescence for Organics \& Chemicals, an instrument used on the Mars Perseverance rover}
\end{marginnote}

There are significant synergies between the geology and biology communities. In particular, the EGS Collab project (Section~\ref{sec:geology}) worked closely with life science researchers. EGS Collab biologists made significant contributions~\cite{EGS-Zhang2025, EGS-Zhang2022, EGS-Zhang2020}, and drill core from the EGS project was also used by a bio-geological team that discovered a set of naturally-occurring extremophiles at SURF (patent pending) that consume CO$_{2}$ gas and convert it into a solid carbonate mineral (MgCO$_{3}$), a process known as carbon mineralization~\cite{Ustunisik2023, GOVIL2024713}.

\subsection{Engineering}

The Thermal Breakout project, led by RESPEC Company LLC, is developing technology using customized heating systems for determining the maximum principal geologic stress magnitude and direction. Several initial heating tests were performed in 2019 and 2020 using existing holes located on the 4850L and 4100L~\cite{10.2118/201195-PA}. Two new 4100L holes were drilled, and heater tests were successfully performed in 2025.

Caterpillar Inc.\ invested in 1700L infrastructure to field test mining equipment and showcase products for customers. After several years, Caterpillar Inc.\ is redirecting resources, leaving developed space available for other uses, as noted in Section~\ref{sec:facilities_ug}.

\subsection{Past Experiments}

Earlier snapshots of the SURF science program are available (for example, see~\cite{Heise:2015vza} and references therein). Notable past SURF projects in addition to those referenced above include the following:

\begin{itemize}
\item Deep Underground Gravity Laboratory (DUGL)~\cite{Coughlin:2019hcf, Meyers2019, Mandic2018, Coughlin:2014yda, Harms:2010mp, Acernese:2010zz}: A subset of the LIGO collaboration performed noise characterization measurements using a 3D seismic array established across seven elevations at SURF in support of gravity-wave research during the time period 2008--2016.
\item Samples collected informally have also resulted in high-impact publications~\cite{Rinke-2014}.
\end{itemize}
\begin{marginnote}[]
\entry{DUGL}{Deep Underground Gravity Laboratory}
\entry{LIGO}{Laser Interferometer Gravitational-Wave Observatory}
\end{marginnote}


\section{COMMUNITY ENGAGEMENT}

Launched in 2020, the SURF User Association~\cite{SURF_UserAssociation} provides an additional framework for communication on topics important to researchers. User Association membership is open to the global underground science community and is managed by a nine-member Executive Committee. As input for U.S.\ strategic planning community meetings in 2022, SURF's User Association coordinated a Long-Term Vision Workshop with discussion and advice on future directions in underground science~\cite{SURF_VisionWorkshop}. A clear message from the workshop was support for additional underground space for all disciplines.

The vision for the Institute for Underground Science at SURF~\cite{SURF_institute} is to foster a globally-recognized intellectual community with impactful programs that engage researchers, educators and students, as well as support underground science initiatives. Formally established in December 2023, initial Institute programs include the CEnter for Theoretical and Underground Physics (* and Related Topics) (CETUP*) workshop and the Quantum Partnerships Workshop as well as a monthly colloquium series~\footnote{The series was renamed in March 2026, recognizing important contributions by John Bahcall.}.
\begin{marginnote}[]
\entry{CETUP*}{CEnter for Theoretical and Underground Physics (* and Related Topics)}
\end{marginnote}


\section{FUTURE PLANS}
\label{sec:future}

SURF's long-term strategic plan is centered on successfully hosting current and future generations of experiments and creating opportunities for new research, including expansion of the facility footprint to exploit SURF's significant potential to meet the needs of the underground science community. Over the past decade, interest in performing scientific studies at SURF has steadily increased -- so far in 2025 there have been expressions of interest from more than 30 research groups, including ongoing interest related to proposals received for SURF's 2024 call for Letters of Interest as well as the wider scientific community.

The underground science community recognizes that the additional space afforded by LBNF (see Table~\ref{tab:footprint}) will significantly enhance progress in neutrino oscillation physics. However, more space is needed to host other high-priority experiments identified in U.S.\ strategic planning for high energy and nuclear physics. To meet that broad set of needs, two SURF levels have been considered for expansion by 2035, namely 4850L and 7400L. Further development of the 4850L affords the most practical opportunity for new space on the timeline of next-generation dark matter and neutrino experiments, while planning for a 7400L laboratory awaits a stronger mandate from the community.

\subsection{4850L Expansion}

\begin{figure}[!htbp]
    \centering
    \includegraphics[width=\textwidth]{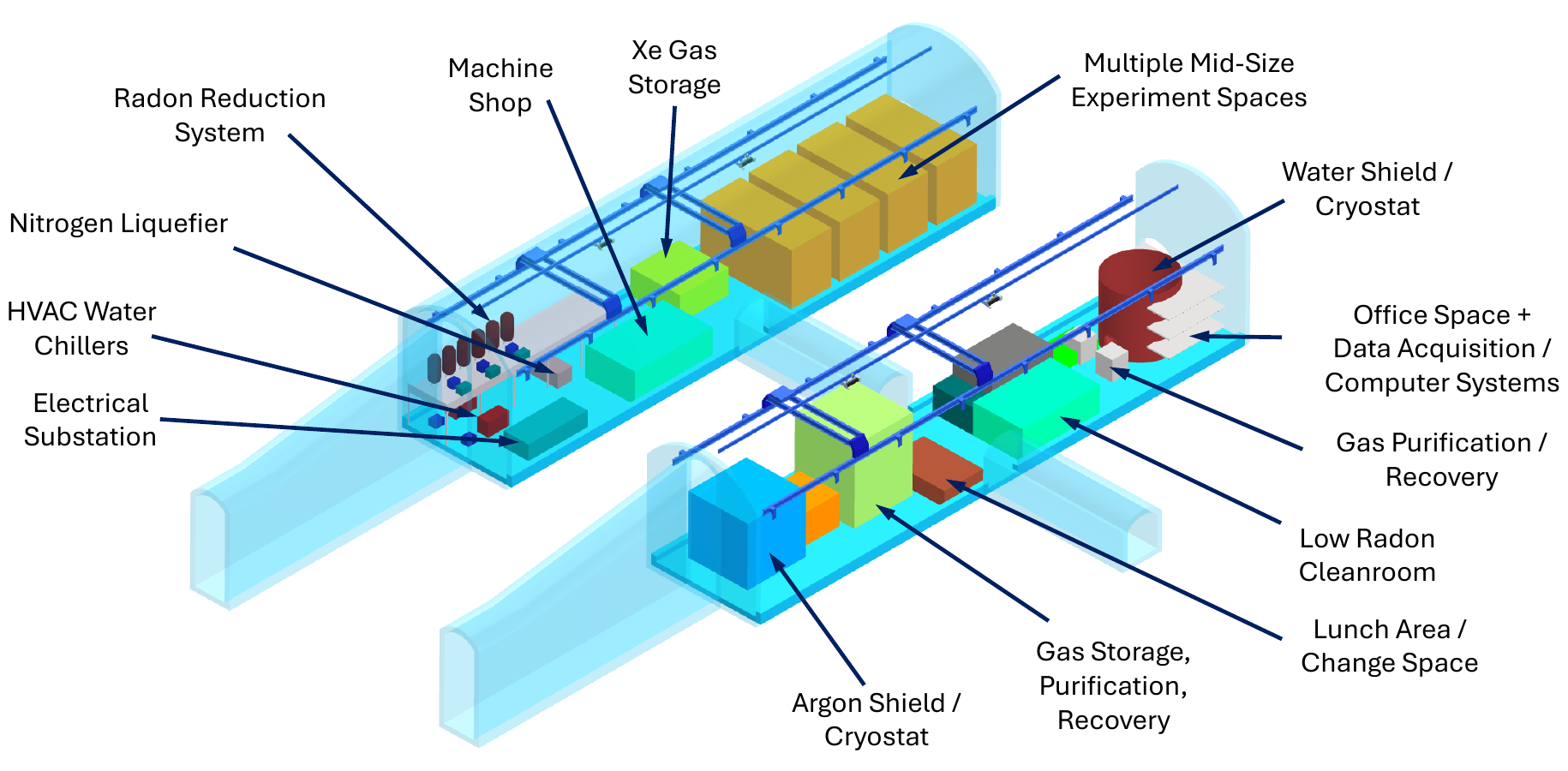}
    \caption{Conceptual drawing of the SURF 4850L Expansion Campus, with up to two 100-m-long caverns (20~m wide $\times$ 24~m high). As shown, the space could allow for 1--2 large experiments as well as several mid-size experiments. Preparations necessary for embarking on this new excavation have been completed.}
    \label{fig:ExpansionLab}
\end{figure}

Preliminary designs exist for up to two new caverns near the Ross Campus to house next-generation experiments requiring larger caverns than currently available in the U.S. With an appropriation of \$13M from the South Dakota legislature in 2023, excavation of the initial expansion phase was completed in March--September 2024. Private funding is being sought to support cavern excavation, aligned with the timeframe for construction of next-generation experiments in the early 2030s. Based on community input (including Snowmass~\cite{Butler:2023eah}), as well as ongoing discussions with collaborations for next-generation experiments, nominal designs for new 4850L caverns accommodate proposed experiments such as a multi-ton-scale 0$\nu\beta\beta$ experiment (e.g., LEGEND-6000~\cite{LEGEND-6000}), one or more third-generation dark matter experiments (e.g., XLZD~\cite{Aalbers:2022dzr}, ARGO~\cite{McDonald:2024osu}), or other detectors that can take advantage of the LBNF neutrino beam to complement DUNE science (e.g., Theia novel liquid scintillator experiment~\cite{Theia-ESPP, Theia-Snowmass, Theia:2019non}); see Figure~\ref{fig:ExpansionLab}.

\subsection{7400L Expansion}

Although SURF's main focus for expansion is the 4850L, several disciplines would benefit from a deep site, including extremophile biology and geothermal projects, but most significantly, physics experiments that are especially sensitive to cosmogenic backgrounds. In particular, future neutrinoless double-beta decay experiments could benefit from being hosted on SURF's 7400L ($\sim$6500~m.w.e.\@), where the cosmic-ray muon flux is expected to be more than 30$\times$ lower than the 4850L, providing the lowest cosmic ray muon flux of any underground laboratory (see Figure~\ref{fig:LabVolumes}) and ensuring cosmogenic backgrounds are negligible. The concept for laboratory space on the 7400L is based on previous studies (1--2 caverns, 15~m $\times$ 15~m $\times$ 75~m), and access would require significant refurbishment.

\subsection{Other New Facilities}

In addition to planning for new underground space to host large-scale, next-generation experiments, existing facilities and future facility renovations may provide opportunities to significantly expand science at SURF.

\subsubsection{Quantum User Facility}
\label{sec:quantum_user_facility}

SURF is exploring opportunities to establish an underground Quantum User Facility at the 4850L that will stimulate new research opportunities exploiting cryogenic calorimeters and quantum phonon sensors for applications in dark matter searches, particle astrophysics enabled by the coherent elastic neutrino-nucleus scattering (CE$\nu$NS) interaction, and neutrinoless double-beta decay searches.

\begin{figure}[!htbp]
    \centering
    \includegraphics[width=0.66\textwidth]{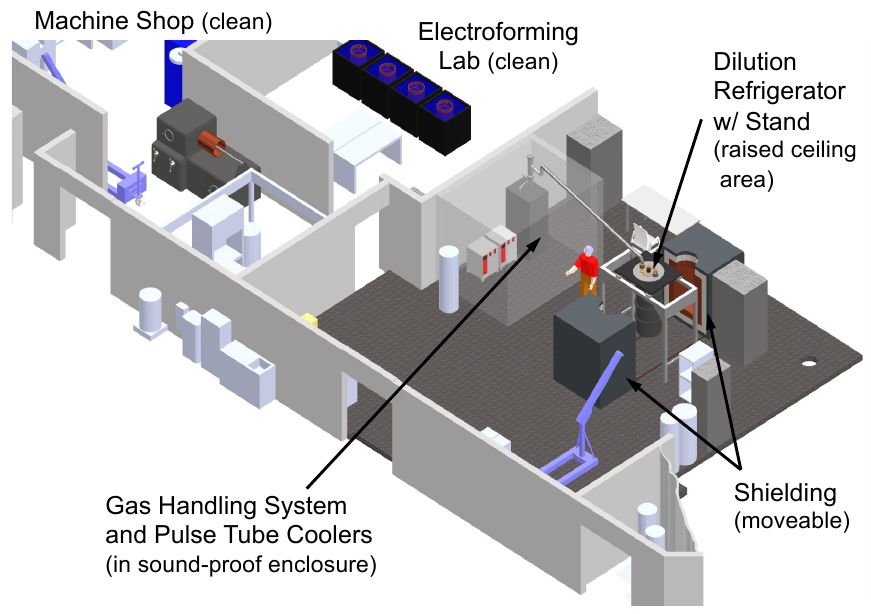}
    \caption{Proposed SURF Quantum User Facility located in a Davis Campus laboratory space that previously supported the {\sc Majorana Demonstrator}. Some facility cooling and electrical modifications will be necessary and could begin in 2026.}
    \label{fig:QuantumUserFacility}
\end{figure}

In 2024, the South Dakota legislature appropriated just over \$3M establishing a Center for Quantum Information Science and Technology involving two state universities. That same year, the SURF Institute launched a new workshop series to enhance possible synergies with state QIS initiatives as well as explore broader interest in topics such as quantum entanglement leveraged in quantum communication networks. Funds are currently being sought to purchase a dilution refrigerator, shielding and perform necessary facility upgrades at the 4850L Davis Campus (see Figure~\ref{fig:QuantumUserFacility}). SURF's Quantum User Facility will bolster the initial state investment with additional scientific staff and support development of novel detectors that will address open questions in fundamental science.

Cryogenic particle detectors work based on the temperature change that occurs when a particle deposits energy within the active volume. These detectors rely on quantum phonon sensors such as heavily-doped semiconductor thermistors, superconducting transition edge sensors (TES) or kinetic inductance detectors (KID); operating temperatures in the $<$50~mK range are required for perceptible temperature changes in response to particle interactions and for the quantum phenomena underpinning the temperature sensor to emerge. Due to the ability to achieve excellent energy resolution, low detector energy thresholds (sub keV) and recent advances in large ultra-low temperature cryogenic systems, these types of detectors are a promising approach for low-mass dark matter searches \cite{SPICE:2023tru, Bass:2023hoi} and neutrinoless double-beta decay searches~\cite{Munster:2017lol}. Since the first observation of CE$\nu$NS, modest-sized arrays of cryogenic calorimeters have been proposed as supernova and reactor neutrino observatories~\cite{COHERENT:2017ipa, RES-NOVAGroupofInterest:2022glt, Ricochet:2023nvt, NUCLEUS:2017htt}.
\begin{marginnote}[]
\entry{TES}{transition edge sensor}
\entry{KID}{kinetic inductance detector}
\end{marginnote}

Current and proposed flagship experiments, such as LZ and XLZD, will explore well-motivated WIMP dark matter over a range of masses (10$^{1}$--10$^{4}$~GeV/c$^{2}$). However, the community has identified a number of new directions that motivate a portfolio of multiple small experiments searching for dark matter over a wide range of masses. These experiments require a low-background, ultra-low temperature test facility that can be accommodated by a deep underground laboratory~\cite{Cardani2021, Vepsalainen:2020trd}. In recent years, several low-mass dark matter experiments have expressed interest in space at SURF using quantum sensors with cryogenic targets such as some of those described in~\cite{Baxter:2025odk}, using narrow-bandgap semiconductors such as Eu$_{5}$In$_{2}$Sb$_{6}$~\cite{Abbamonte:2025guf}, or using germanium-based quantum systems~\cite{Mei:2025ozs, Mei:2025evp}.

\subsubsection{Vertical Facility}

For decades, there has been significant interest in the potential science associated with a vertical facility. During early DUSEL planning, workshops highlighted interest in atom interferometry as well as cloud formation and microgravity studies. During the Snowmass 2021 community planning effort, SURF was identified as a possible site for future gravitational-wave detectors based on atom interferometry~\cite{Ballmer:2022uxx}.

Atom interferometry is a promising technique based on the superposition and interference of atomic wave packets for probing aspects of fundamental physics, astrophysics and cosmology, including ultralight dark matter (mass range $\sim$10$^{-19}$--10$^{-11}$~eV, complementary to direct and other searches) and gravitational waves (signals in frequency range 0.1--10~Hz, characteristic of mergers of intermediate-mass black holes (10$^{2}$--10$^{5}$~M$_{\odot}$), complementary to laser interferometry searches).

\begin{figure}[!htbp]
    \centering
    \includegraphics[width=\textwidth]{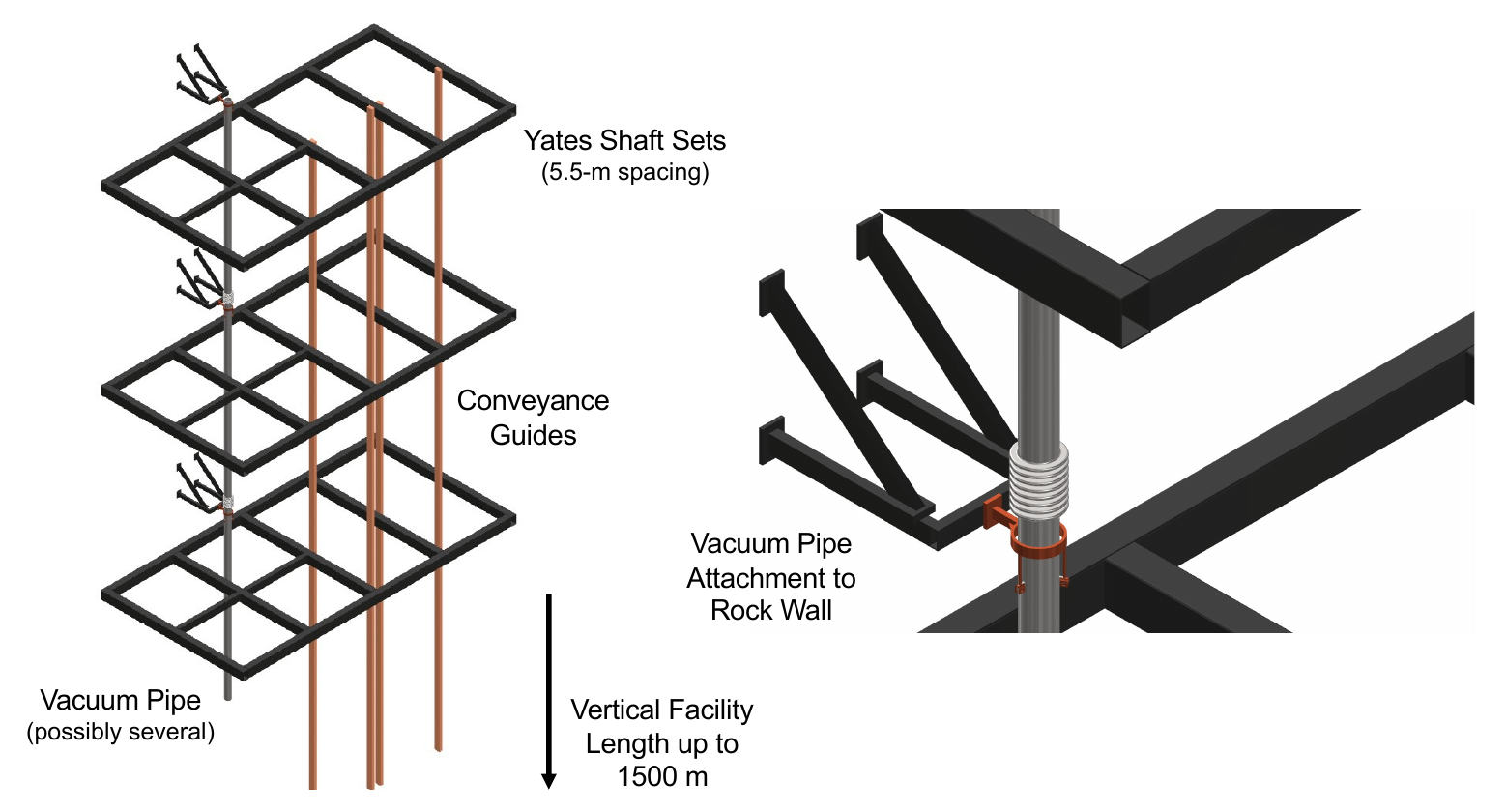}
    \caption{Proposed SURF Vertical Facility up to 1500~m in length and possible options for multiple pipes. Rock wall attachments are based on the design for other piping in the Ross Shaft. Development can be accommodated in conjunction with a planned Yates Shaft rehabilitation in the 2030s.}
   \label{fig:VerticalFacility}
\end{figure}

Following discussions at the 2021 SURF Vision Workshop, a survey was completed in 2022 that identified six legacy shafts as potentially feasible (but challenging) for a vertical facility. A planned upgrade of one of the Yates Shaft, necessary to ensure long-term safe and redundant access at SURF, may afford an opportunity to allocate one compartment of the structure to such a unique facility (nominally, 1.75~m $\times$ 2.0~m); see Figure~\ref{fig:VerticalFacility}. The renovation work is expected to take place in the 2030s, after related upgrades are complete (electrical, hoists). Relatively close proximity to the planned Qunatum User Facility at the 4850L Davis Campus could have significant benefits.

In October 2025, SURF joined the Terrestrial Very-Long-Baseline Atom Interferometry (TVLBAI) proto-collaboration that seeks to advance planning in this area~\cite{TVLBAIProto:2025wyn, Abdalla:2024sst, Abend:2023jxv}. A 100-m vertical facility is being constructed at Fermilab for the MAGIS-100 experiment~\cite{MAGIS-100:2021etm}, with opportunities for partnership.
\begin{marginnote}[]
\entry{TVLBAI}{Terrestrial Very-Long-Baseline Atom Interferometry}
\end{marginnote}


\section{SUMMARY}

SURF is a deep underground research facility dedicated to scientific uses that has been operating for more than 18~years, offering world-class service and a proven track record of enabling experiments to deliver high-impact science across diverse scientific communities. Research activities, including dark matter, neutrino and nuclear astrophysics, are supported at surface and underground facilities. Significant work is also being performed at SURF by researchers in geology, biology and engineering topics. Electroformed copper production and assay capabilities are also available. 

In addition to the existing science program as well as hosting LBNF/DUNE, SURF is eager to host future experiments, and to that end, expansion at SURF is on the horizon. In addition to the existing 4850L campuses, SURF has already started to increase underground laboratory space on the 4850L. A mixture of federal, state and private funding will allow phased development of underground space aligned with needs for next-generation neutrino and dark matter projects. Other facilities actively being investigated at SURF include a Quantum User Facility and a Vertical Facility, both of which would support quantum information science, including searches in expanded parameter spaces for low- and ultra-low-mass dark matter as well as gravitational waves.





\section*{DISCLOSURE STATEMENT}
The author is not aware of any affiliations, memberships, funding, or financial holdings that might be perceived as affecting the objectivity of this review. 

\section*{ACKNOWLEDGMENTS}
The author wishes to recognize the large number of scientific colleagues and technical staff over many years who contributed to the realization of the facility and science programs reported in this review, with special regards to Kevin Lesko for his early vision and perseverance. In addition, the author acknowledges helpful discussions with Bill Roggenthen (site geology) and Vitaly Kudryavtsev (calculations of rock atomic properties). SURF construction was financed by the State of South Dakota as well as philanthropist T.\ Denny Sanford. Current operation of SURF is sponsored by the U.S.\ Department of Energy, Office of Science, Office of High Energy Physics under Award Number DE-SC0020216. The author further acknowledges the South Dakota Governor's office, the South Dakota Community Foundation, the South Dakota State University Foundation, and the University of South Dakota Foundation for investments that facilitated the purchase of xenon. Finally, SURF respectfully acknowledges that the facility is located on the traditional land of Indigenous American peoples and honors their rich cultural heritage and enduring contributions.


\bibliographystyle{ar-style5} 
\bibliography{main}
 
\end{document}